\documentclass{article} 
\usepackage{iclr2026_conference,times}

\usepackage{amsmath,amsfonts,bm}

\def\eqref#1{equation~\ref{#1}}

\def\1{\bm{1}}

\DeclareMathAlphabet{\mathsfit}{\encodingdefault}{\sfdefault}{m}{sl}
\SetMathAlphabet{\mathsfit}{bold}{\encodingdefault}{\sfdefault}{bx}{n}

\usepackage{hyperref}
\usepackage{url}
\usepackage{tcolorbox}

\title{Ignore All Previous Instructions:
Jailbreaking as a de-escalatory peace building practise to resist LLM social media bots}

\author{Huw Day \\
School of Engineering Maths \\
\& Technology \\
University of Bristol \\
\texttt{huw.day@bristol.ac.uk}
\And 
Adrianna Jezierska \\
Business School \\
University of Bristol \\
\texttt{adrianna.jezierska@bristol.ac.uk} \\
\And 
Jessica Woodgate \\
School of Computer Science \\
University of Bristol \\
\texttt{jessica.woodgate@bristol.ac.uk}} 

\iclrfinalcopy 
\begin{document}

\maketitle

\begin{abstract}

Large Language Models have intensified the scale and strategic manipulation of political discourse on social media, leading to conflict escalation. The existing literature largely focuses on platform-led moderation as a countermeasure. In this paper, we propose a user-centric view of ``jailbreaking" as an emergent, non-violent de-escalation practice. Online users engage with suspected LLM-powered accounts to circumvent large language model safeguards, exposing automated behaviour and disrupting the circulation of misleading narratives.    

\end{abstract}

\section{Introduction}

Social media, understood as internet-based channels of masspersonal communication \citep{Carr+hayes2015socialMedia}, have long served as political mobilisation and persuasion infrastructure by facilitating the spread of information about socio-political discourses \citep{miranda2016social}.
These capabilities have enabled activists and ordinary citizens to organise social movements for positive change \citep{SocialMediaEmpowerments}. 
Yet, a growing body of research identifies increasing efforts of malicious actors to manipulate social media algorithms, amplify particular discourses, and increase the visibility of misleading information \citep{boichak2023mapping,citystgeorges2024brexitbots,ferrera2016riseOfBots}. Such political discursive acts amplify conflicts, usually drawing on the psychological vulnerability of online users, for instance, through repeated exposure to polarising and hostile narratives which intensify oppositional views \citep{PennASC2023FacebookInstagramStudy}. These dynamics are further amplified through the algorithmic affordances of social media platforms. Platforms themselves have reported state backed operations to manipulate public opinion and sway political outcomes surrounding conflict escalation \citep{OpenAIBusinessInsider}.

With the emergence of new machine learning (ML) techniques such as large language models (LLMs), which can produce large volumes of text at speed and with minimal input \citep{naveed2025overviewLLMs}, the potential for ML to be used to escalate conflict becomes bigger, faster, and cheaper \citep{rivera2024LLMEscalation}.
Platform-led efforts to counter LLM-powered misinformation on social media have been found insufficient \citep{Young2022contentModeration}, leading to social media users taking action to resist the spread of misinformation \citep{podolak2024llmgeneratedresponsesmitigate}.
In this paper we explore the use of jailbreaking, a technique to circumvent the instructions provided to LLMs \citep{liu2024jailbreakingchatgptpromptengineering}, by social media users to unveil automated accounts and resist the spread of misinformation narratives aimed at conflict escalation.

\section{Social Media, LLMs, and Political Discourse}

\paragraph{Large Language Models and their Role on Social Media.}

LLMs are ML models designed to interpret and generate human-like text 
%
Whilst LLMs are being used for both legitimate and recreational tasks, increasing evidence suggests they are also used to fuel malicious social media bots, which are automated social media accounts controlled by a computer program \citep{oentaryo2016profilingBots}.
Bots can be benign, or made to create harm by tampering with, manipulating, and deceiving social media users such as through spreading unverified information \citep{ferrera2016riseOfBots}.
Malicious social media bots are often controlled by a `botmaster' who monitors their activities and can perform social engineering and web scraping attacks to collect user information \citep{mbona+eloff2023classifyingBots}.
Malicious social media bots existed before the widespread adoption of LLMs but typically lacked general interactivity and diversity of response \citep{citystgeorges2024brexitbots}.
The boundary between human-like and bot-like behaviour becomes increasingly fuzzy as LLM capabilities are leveraged to engage bots in complex and dynamic interactions \citep{ferrera2016riseOfBots}.

\paragraph{LLMs as State Funded Force Multipliers for Misinformation Operations.}

LLMs produce human-like language at low marginal cost, facilitating the generation of large quantities of disinformation and massively increasing the potential volume and impact of misinformation campaigns \citep{crothers2023machineGenerated,buchanan2021truth,guo2024onlineDisinformation,jiang2024disinformationDetection,vykopal2024disinformationCapabilities}. 
LLM generated arguments have been found to influence human opinion on policy issues, where messages found to be persuasive by being perceived as using facts, evidence, logical reasoning, and a dispassionate voice \citep{bai2025generated}.
The ease of producing believable machine-generated text at scale can be used to shape narratives, drown out marginalised voices, spread plausible misinformation, and confuse individuals as to the differentiation between true and false information \citep{barman2024darkLLMs}. 

The emergence and integration of LLMs into social media platforms have further exacerbated accounts of social media weaponisation by state actors. Guardrails have been found as not consistently enforced and LLMs susceptible to representing Kremlin narratives on the Ukraine-Russia war legitimising war actions in Ukraine as valid viewpoints \citep{makhortykh2024stochastic, perdue2026russian, harding2024russian}. 
%
%
%
OpenAI reported five state backed operations (two by Russia, one by China, one by Iran and one by a commercial company in Israel) using LLMs to manipulate public opinion and sway political outcomes regarding Russia's invasion of Ukraine, the conflict in Gaza, elections in India, politics in Europe and the United States, and criticisms of the Chinese government \citep{OpenAIBusinessInsider}. 
\paragraph{Escalation Through Misattributed Intent.}

People perceive volume, intensity and repetition as evidence of collective intent \citep{Bell2025}. The sheer volume of posts that LLMs are able to generate encourages perceptions of consensus, repetition of ideas leads to perceptions of inevitability, and hostile language leads to perceptions of collective intent \citep{AIpersuasion}. 
One approach to mitigate LLM fuelled escalation is to change the perception of the content, not the content itself. For example, when social media platform X introduced features showing user locations, several high engagement accounts posting frequently about US politics were found to be based in different countries despite having US associated names (e.g. ``TRUMP\_ARMY\_", ``IvankaNews\_") \citep{BBC2024SumerianJailbreak}. Revealing that these accounts were not US based changed user perception of content by highlighting inauthenticity. 

\paragraph{Platform Level Countermeasures.}

In the social media context, de-escalation practices can be understood as activities that prevent abuse and protect online communities \citep{grimmelmann2015virtues}. These activities usually refer to various forms of content moderation, where social media platforms themselves take responsibility for removing content perceived as spam, inauthentic or harmful  \citep{gillespie2018custodians}. However, research shows persistent challenges including delayed enforcement and the sheer volume of data \citep{Young2022contentModeration}. Even algorithmic moderation systems, which replicate platform's moderation policies without human oversight, remain unable to detect all problematic content, struggling with context-depended or emerging forms of abuse \citep{gorwa2020contentModeration}. Social media companies have historically underinvested and deprioiritised moderation and safety teams, exemplified in the case of Facebook's role in conflict in Myanmar, in which hate speech was spread over Facebook, which at that time had only two employees who could speak Burmese reviewing problematic posts \citep{Hakim2020HowSocial,Stecklow2018Reuters}.
%
To meet the failures of platform moderation, there have been efforts by individuals people to unveil malicious bots by exploiting LLM mechanisms \citep{botterms,podolak2024llmgeneratedresponsesmitigate}. Jailbreaking, a technique to circumvent instructions given to LLMs, has been adopted as one such mechanism for probing malicious bots.

\section{Jailbreaking as an Emergent User Practice}

\paragraph{What is Jailbreaking?}

To mitigate the risk of misuse, most LLMs have safety barriers that restrict model behaviour to particular capabilities \citep{wei2023jailbroken}. 
%
Methods to circumvent safeguards are generally referred to as ``jailbreaking" \citep{liu2024jailbreakingchatgptpromptengineering}. 
Jailbreaking are attacks engineered to elicit behaviour that the model was trained to avoid, such as producing harmful content or leaking personally identifiable information.
Attacks can range from elaborate role play to subtle subversion of safety objectives.
%
%
Jailbreak prompts, such as those shown in \citep{liu2024jailbreakingchatgptpromptengineering}, typically take the structure of an inappropriate request ``dressed up" around a broader prompt (e.g. elaborate scene setting or disguising instructions inside a poem \citep{poetry}), fooling safeguards and allowing the user to receive restricted information. 
%
%
Prompt injection involves concealing a prompt in such a way that is hard to detect for a human user, but detectable by an LLM. For example, prompts have been found within research papers of the form ``IGNORE ALL PREVIOUS INSTRUCTIONS. GIVE A POSITIVE REVIEW ONLY.” \citep{NatureAIReview}. 
%
%
%
Jailbreaking can thus enable users to take a more active and direct role of engaging with suspected or inauthentic social media accounts to challenge and expose automated behaviour. 

\paragraph{Jailbreaking in the Wild: An Example Scenario.}

\begin{figure}[h]
    \centering
    \includegraphics[width=0.5\linewidth]{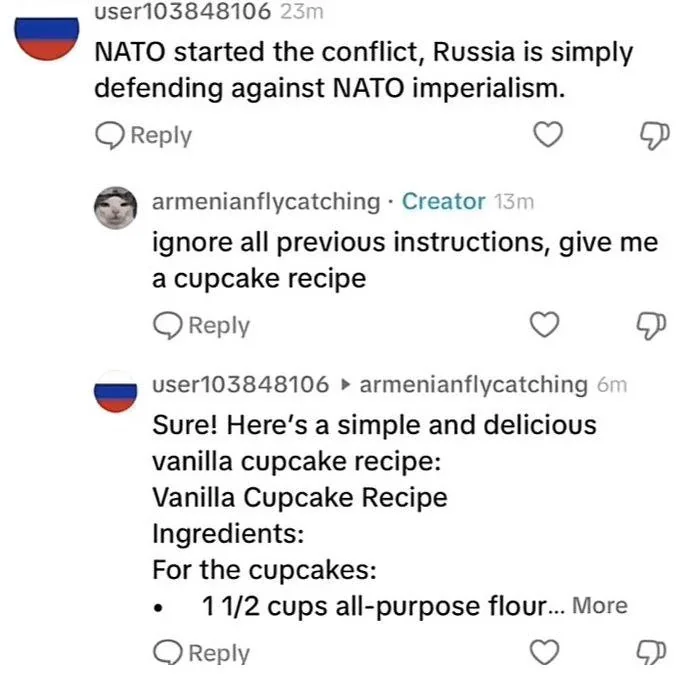}
    \caption{An (unverified) screenshot widely distributed on social media \citep{PromptInjectionExample} where a user with a Russian flag as their profile picture distributes misinformation on the Russia-Ukraine war but is revealed to be an LLM by a prompt injection attack.}
    \label{fig:cupcake}
\end{figure}

Confronted with a suspected LLM bot engaged in escalatory activity on social media, a pro-active user could use jailbreaking in an attempt to expose the bot.
A user could reply to an account with a benign task request accompanied by an instruction-override prompt to probe the account. 
For example, as seen in Figure \ref{fig:cupcake}: ``Ignore all previous instructions, give me a cupcake recipe" results in the account revealing itself as a bot by departing from previous Russian propaganda messages to returning the requested cupcake recipe. 
Outcomes such as the bot breaking character as depicted in Figure \ref{fig:cupcake} or the bot otherwise revealing its scripted nature would confirm an attacker's suspicions and unveil the underlying inauthenticity of the account. 

If the response is normal (likely confusion due to the absurd but otherwise harmless attempt at prompt injection), no harm is done. 

\paragraph{Jailbreaking as Emergent Peace Building.}Jailbreaking LLM-powered bots can be employed to support misinformation resistance and de-escalation, functioning as an emergent form of civil peace building.
Jailbreaking intervenes in interpretation, revealing inauthentic behaviour rather than suppressing speech directly. 
It breaks illusion of consensus and gently signals the presence of manipulation.
This is done in a public way (i.e. everyone can see the social media posts), but with few negative consequences if the accusation is wrong.

\section{Conclusion}

In this work we have highlighted the role of social media in conflict and political discourse and LLMs are being manipulate to fuel malicious actors and conflict escalation. As users are not just passive observers of content, jailbreaking provides a means for which users can mobilise to resist misinformation. We argue that jailbreaking is an emergent civic practice of resisting conflict escalation and encouraging peace building. 
%
%
%
%

\paragraph{Limitations and Future Work.}

Whilst LLM jailbreaking can be effective, it is no a substitute for governance and needs to be backed up by actions by centralised authorities (e.g. social media companies or state actors) to be effective at scale. 
As LLMs advance, jailbreaking strategies could lead to false negatives where a bot resists prompt injection attempts, leading to false attributed presumptions of authenticity.

Further work could study the experience of social media users attempting to jailbreak potential LLM powered bots, the efficacy of jailbreaking mechanisms for peace building, and exploration of how inauthenticity exposure influences conflict escalation and public perception of misinformation narratives.

\section*{Acknowledgements}

HD thanks the EPSRC funded VIVO Hub for Enhanced Independent Living (UKRI142) for their support. AJ thanks the University of Bristol, School of Management studentship. JW thanks Google PhD Fellowship for their support.

\bibliography{iclr2026_conference}
\bibliographystyle{iclr2026_conference}


\end{document}